\documentclass[prd,preprint,showpacs,preprintnumbers,amsmath,amssymb,eqsecnum]{revtex4-1}

\usepackage{graphicx}
\usepackage{dcolumn}
\usepackage{bm}
\usepackage{subfigure}
\usepackage{color}

\begin{document}

\preprint{RUP-14-17}

\title{Negative tension branes as stable thin shell wormholes}

\author{Takafumi Kokubu}
 \email{takafumi@rikkyo.ac.jp}
\author{Tomohiro Harada}%
 \email{harada@rikkyo.ac.jp}
\affiliation{%
Department of Physics, Rikkyo University, Toshima, Tokyo 171-8501,
Japan} 
\date{\today}

\begin{abstract}
We investigate negative tension branes as stable thin shell 
wormholes in 
Reissner-Nordstr\"{o}m-(anti) de Sitter spacetimes
in $d$ dimensional Einstein gravity.
Imposing Z2 symmetry, we construct and classify 
traversable static thin shell wormholes 
in spherical, planar (or cylindrical) and hyperbolic symmetries.
In spherical geometry, we find the 
 higher dimensional counterpart of
 Barcel\'o and Visser's wormholes, which are stable against 
 spherically symmetric perturbations. 
We also find the classes of thin shell wormholes in planar and hyperbolic
symmetries with a negative cosmological constant, which 
are stable against perturbations preserving symmetries. 
In most cases, stable wormholes are found 
with the combination of 
an electric charge and a negative cosmological constant.
However, as special cases, we find stable wormholes even 
with vanishing cosmological constant in spherical symmetry 
and with vanishing electric charge in hyperbolic symmetry.
\end{abstract}

\maketitle

\newpage

\tableofcontents

\newpage

\section{Introduction}
Wormholes are spacetime structures which connect two different universes or two points of our universe.
M.~Morris and K.~Thorne have pioneered qualitative study for static
spherically symmetric wormholes which are two-way traversable and
discussed the traversable conditions in~\cite{morris&thorne}. The most
difficult requirement  for satisfying the traversable conditions is an
exotic matter which violates the null or weak energy condition. In
general relativity, Morris and Thorne's static spherically symmetric
traversable wormholes need the stress-energy tensor that violates 
the energy condition at their throat. See details in~\cite{morris&thorne}.

An another class of traversable wormholes has been found by
M.~Visser. This class of wormholes can be obtained by a
``cut-and-paste'' procedure~\cite{visser} and such structures are called
{\it thin shell} wormholes (TSWs). E.~Poisson and M.~Visser first
presented stability analysis for spherical perturbations
around such TSWs, and found that there are stable configurations
according to the equation of state of an exotic matter residing on the
throats~\cite{poisson&visser}. The work of Poisson and Visser has been
extended in different directions; charged TSWs~\cite{Reissner}, TSWs
constructed by a couple of Schwarzschild spacetimes of different masses
\cite{Ishak&Lake} and TSWs with a cosmological constant
\cite{cosmologicalconstant}. TSWs in cylindrically symmetric spacetimes
have also been studied~\cite{cylindricalTSW}. Garcia {\it et al.}
published a paper about stability for generic static and spherically
symmetric TSWs~\cite{Garcia+}. Dias and Lemos studied stability in
higher dimensional Einstein gravity~\cite{Dias&Lemos}.

Pure tension branes, whether the tension is positive or negative, have
particular interest because they have Lorentz invariance and have no
intrinsic dynamical degrees of freedom. In the context of stability, 
pure negative tension branes have no intrinsic instability by their own,
although they violate the weak energy condition. 
This is in contrast with the Ellis wormholes, for which a phantom scalar
field is assumed as a matter content and it suffers the so-called ghost
instability because of the kinetic term of a wrong sign~\cite{scalarfield,Ellis instability}. 
The construction of traversable wormholes by using negative tension
branes have first been proposed by C.~Barcel\'o and M.~Visser
\cite{barcelo&visser}. They analyzed dynamics of spherically symmetric
traversable wormholes obtained by operating the cut-and-paste procedure
for negative tension 2-branes (three dimensional timelike singular 
hypersurface) in 
four dimensional spacetimes. 
They found stable brane wormholes constructed by pasting a couple of
Reissner-Nordstr\"om-(anti) de Sitter spacetimes. In their work
\cite{barcelo&visser}, the charge is essential to sustain such
wormholes. And in most cases, a negative cosmological constant tends to
make the black hole horizons smaller. However, in exceptional cases, one
can obtain wormholes with a vanishing cosmological constant, if 
the absolute value of the charge satisfies a certain condition.

In this paper, we investigate negative tension branes as thin shell
wormholes in spherical, planar (or cylindrical) and hyperbolic
symmetries in $d$ dimensional Einstein gravity with an
electromagnetic field and a cosmological constant in bulk
spacetimes. In spherical geometry, we find the higher dimensional counterpart of Barcel\'o and Visser's wormholes which are stable against spherically symmetric perturbations. As the number of dimensions increases, larger charge is allowed to construct such stable wormholes without a cosmological constant. Not only in spherical geometry, but also in planar and hyperbolic geometries, we find static wormholes which are stable against perturbations preserving symmetries. 

This paper consists of the following sections. In Sec. II, we present a
formalism for wormholes, which is more general than previous formalisms
and also obtain a stability condition against perturbations
preserving symmetries. In Sec. III, we introduce wormholes with a
negative tension brane and we analyze the existence of static solutions,
stability and horizon avoidance in spherical, planar and hyperbolic symmetries. Sec. IV is devoted to summary and discussion.

\section{Wormhole formalism}

\subsection{Construction}
The formalism for spherically symmetric $d$ dimensional thin shell
wormholes has been developed first by Dias and Lemos
\cite{Dias&Lemos}. We extend their formalism to more general
situations. We obtain wormholes by operating three steps invoking
junction conditions~\cite{toolkit}. This approach to construct TSWs 
has been pioneered by Poisson and Visser~\cite{poisson&visser}.

Firstly, consider a couple of $d$ dimensional manifolds, $\mathcal
V_\pm$. We assume $d\ge 3$.
The $d$ dimensional Einstein equations are given by
\begin{eqnarray}
G_{\mu\nu \pm}+\frac{(d-1)(d-2)}{6}\Lambda_\pm g_{\mu\nu \pm}=8\pi T_{\mu\nu \pm},
\end{eqnarray}
where $G_{\mu\nu \pm}$,  $T_{\mu\nu \pm}$ and $\Lambda_\pm$ are Einstein
tensors, stress-energy tensors and cosmological constants in the
manifolds $\mathcal V_{\pm}$, respectively. 
The metrics on $\mathcal V_\pm$ are given by $g^\pm_{\mu\nu}(x^\pm)$. 
The metrics for static and spherically, planar and hyperbolically
symmetric spacetimes  on $\mathcal V_\pm$ are written as 
\begin{eqnarray}
ds_\pm^2&=&-f_\pm(r_\pm)dt^2_\pm+
f_\pm(r_\pm)^{-1}dr^2_\pm+r^2_\pm (d\Omega_{d-2}^k)^2_\pm ,\label{senso}\\
f_\pm(r_\pm)&=&k-\frac{b_\pm(r_\pm)}{r_\pm}:=k-\frac{\Lambda_\pm r_\pm^2}{3}-\frac{M_\pm}{r_\pm^{d-3}}+\frac{Q_\pm^2}{r_\pm^{2(d-3)}}, \label{f}
\end{eqnarray}
respectively. 
$M_\pm$ and $Q_\pm$ correspond to the masses and charges in $\mathcal
V_\pm$, respectively. $k$ is a constant that determines the 
geometry of the $(d-2)$ dimensional space and takes $\pm 1$ or $0$. 
$k=+1$, $0$ and $-1$ correspond to a sphere, plane (or cylinder) 
and a hyperboloid, respectively. $(d\Omega_{d-2}^k)^2$ is given by
\begin{equation}
(d\Omega_{d-2}^k)^2 =     \begin{cases}
 					k =1 :& (d\Omega^1_{d-2})^2 = d\theta_1^2 + \sin^2 \theta_1d\theta_ 2^2 + \ldots + \prod_{i=2}^{d-3}\sin^2 \theta_id\theta_{d -2}^2 \\
					k =0 :& (d\Omega^0_{d - 2})^2 = d\theta_1^2 + d\theta_2^2 + \ldots + d\theta_{d - 2}^2 \\
					k =-1 :& (d\Omega^{-1}_{d - 2})^2 = d\theta_1^2 + \sinh^2 \theta_1d\theta_2^2 + \ldots + \sinh^2 \theta_1\prod_{i = 2}^{d-3}\sin^2\theta_id\theta_{d-2}^2.
					\end{cases}
\end{equation}

We should note that by generalized Birkhoff's
theorem~\cite{sato-kodama}, the metric (\ref{senso}) is the unique
solution of Einstein equations of electrovacuum for $k=\pm 1$. However,
this is {\it not} unique for $k=0$. Therefore, we should regard
Eq.~(\ref{senso}) as a special electrovacuum spacetime for $k=0$. 

Secondly, we construct a manifold $\mathcal V$ by gluing $\mathcal
V_\pm$ at their boundaries. We choose the boundary hypersurfaces 
$\partial \mathcal V_\pm$ as follows:
\begin{eqnarray}
\partial \mathcal V_\pm \equiv \{r_\pm =a\ |\ f_\pm(a)>0 \},
\end{eqnarray}
where $a$ is called the thin shell radius.
Then, by gluing the two regions $\tilde{\mathcal V}_\pm$ which are defined as
\begin{eqnarray}
\tilde{\mathcal V}_\pm \equiv \{r_\pm  \geq a\ |\ f_\pm(a)>0 \}
\end{eqnarray}
with matching their boundaries, $\partial \mathcal V_+=\partial \mathcal
V_- \equiv \partial \mathcal V$, we can construct a new manifold 
$\mathcal V$ which has a  wormhole throat at $\partial \mathcal
V$. $\partial \mathcal V$ should be a timelike hypersurface, on which the line element is given by
\begin{eqnarray}
ds_{\partial \mathcal V}^2=-d\tau^2+a(\tau)^2(d\Omega_{d-2}^k)^2.
\end{eqnarray}
It is convenient to define the function $F(x)\equiv r-a(\tau)$ so that 
$\partial \mathcal V$ is given by 
\begin{eqnarray}
F=r-a(\tau)=0.  \label{F}
\end{eqnarray}
$\tau$ stand for proper time on the junction surface $\partial \mathcal V$ whose position is described by the coordinates $x^\mu(\xi^i)=x^\mu(\tau,\theta_1,\theta_2,\ldots,\theta_{d-2})=(t(\tau),a(\tau),\theta_1,\theta_2,\ldots,\theta_{d-2})$, where Greek indices run over $1,2,\ldots,d$ and Latin indices run over $1,2,\ldots,d-1$. $\{\xi^i\}$ are the intrinsic coordinates on $\partial \mathcal V$.

Thirdly, by using the junction conditions, we derive the
equations for the manifold $\partial \mathcal V$. To achieve this, we
define unit normals to hypersurfaces $\partial \mathcal V_\pm$. The unit normals are defined by
\begin{eqnarray}
n_{\alpha \pm} \equiv \pm \frac{F_{,\alpha}}{|g^{\mu\nu}F_{,\mu}F_{,\nu} |^\frac{1}{2}}. \label{unitnormal}
\end{eqnarray}
To construct thin shell wormholes, we make the unit normals on $\partial
\mathcal V_\pm$ take different signs, while to construct normal thin
shell models, the unit normals are chosen to be of same signs.
We define vectors tangent to $\partial \mathcal V$ as
\begin{eqnarray}
e_{(i)\pm}^\alpha \equiv \frac{\partial x_\pm^\alpha}{\partial \xi^i}.
\end{eqnarray}
We also define the four-velocity $u_\pm^\alpha$ of the boundary as
\begin{eqnarray}
u_\pm^\alpha    &=&  e_{(\tau)\pm}^\alpha=  (\dot{t}_\pm,\dot{a},0,\ldots,0) \nonumber \\
			&=& \left(\frac{1}{f_\pm(a)}\sqrt{f_\pm(a)+\dot{a}^2}, \dot{a},0,\ldots,0\right),
\end{eqnarray}
where $\dot{} \equiv \partial/\partial \tau$ and
$u^{\alpha}u_{\alpha}=-1$ is satisfied. From Eqs. (\ref{F}) and (\ref{unitnormal}) we have
\begin{eqnarray}
n_{\alpha \pm}=\pm\left(-\dot{a},\frac{\sqrt{f_\pm+\dot{a}^2}}{f_\pm},0,\ldots,0\right)
\end{eqnarray}
and the unit normal satisfies $n_\alpha n^\alpha=1$ and $u^\alpha n_\alpha=0$.

\subsection{Equations for the shell}
In general, there is matter distribution on $\partial \mathcal V$. The
equations for the shell are given by the junction conditions of the 
Einstein equations~\cite{toolkit} as follows:
\begin{eqnarray}
&&S^i_j=-\frac{1}{8\pi}(\kappa^i_j-\delta^i_j\kappa^l_l),  \label{S^i_j} \\
&&\kappa^i_j=\left.(K^{i +}_j-K^{i -}_j)\right|_{\partial \mathcal V},
\end{eqnarray}
where $S^i_j$ is the surface stress-energy tensor residing on $\partial \mathcal V$.
$K_{ij}^\pm$ are the extrinsic curvatures defined by
\begin{eqnarray}
K_{ij}^\pm \equiv (\nabla_\mu n_\nu^\pm) e_{(i)\pm}^\mu e_{(j)\pm}^\nu.  \label{K_ij}
\end{eqnarray}
The non zero components of $K_{i j}^\pm$ are the following:
\begin{eqnarray}
&&K_\tau^{\tau \pm}=\pm (f_\pm +\dot{a}^2)^{-\frac{1}{2}}(\ddot{a}+\frac{1}{2}f_\pm^\prime), \\
&&K_{\theta_1}^{\theta_1 \pm}=K_{\theta_2}^{\theta_2 \pm}=\ldots=K_{\theta_{d-2}}^{\theta_{d-2} \pm}=\pm \frac{1}{a}\sqrt{f_\pm+\dot{a}^2},
\end{eqnarray}
where $' \equiv d/d a$. Consequently, the non zero components of $\kappa^i_j$ are
\begin{eqnarray}
&&\kappa^\tau_\tau=\frac {B_+(a)} {A_ +(a)} + \frac {B_-(a)} {A_-(a)} ,\\
&&\kappa_{\theta_1}^{\theta_1}=\ldots=\kappa_{\theta_{d-2}}^{\theta_{d-2}}=\frac{A_+(a)+A_-(a)}{a},
\end{eqnarray}
where
\begin{eqnarray}
A_\pm(a) \equiv \sqrt{f_\pm+\dot{a}^2} \ ,\ B_\pm(a) \equiv \ddot{a}+\frac{1}{2}f_\pm^\prime.
\end{eqnarray}

Since our metrics (\ref{senso}) are diagonal, $S^i_j$ is also diagonalized and written as
\begin{equation}
S^i_j={\rm diag}(-\sigma,p,p,\ldots,p), \label{s,p,p} \\
\end{equation}
where $p$ is the surface pressure (of opposite sign to surface tension) and $\sigma$ is the surface energy density living on the thin shell.
From Eqs. (\ref{S^i_j}) and (\ref{s,p,p}), we obtain 
\begin{eqnarray}
&& \sigma = -\frac {d - 2}{8\pi a} (A_++ A_ -),  \label{energy density} \\
&& p = \frac{1}{8\pi} \left[ \frac{B_ +}{A_ +} + \frac {B_-} {A_-} + \frac{d - 3}{a}(A_++ A_-)\right]. \label{surface pressure}
\end{eqnarray}
Thus, we deduce a critical property of wormholes that $\sigma$ must be negative.

The equation of motion for the surface stress-energy tensor $S^i_j$ is given by
\begin{eqnarray}
S^{ij}_{\ \ |j}+[T^\alpha_\beta e^{i}_\alpha n^\beta]_{\pm}=0 \label{eom for layer}
\end{eqnarray}
as in~\cite{gravitation} , where ``$[\ ]_{\pm}$'' denotes the difference
between the quantities in ${\mathcal V}_{+}$ and in ${\mathcal V}_{-}$. Since the stress-energy tensor $T^\alpha_\beta$ in the bulk spacetime only contains the electromagnetic field,
\begin{eqnarray}
T^\alpha_\beta=\frac{Q^2}{8\pi r^4} {\rm diag} (-1,-1,1,1),
\end{eqnarray}
one can find $T^\alpha_\beta e^{i}_\alpha n^\beta=0$. Hence Eq.~(\ref{eom for layer}) yields 
\begin{equation}
\frac{d}{d\tau}(\sigma a^{d-2})+p\frac{d}{d\tau}(a^{d-2})=0. \label{conservationlaw}
\end{equation}
Eq.~(\ref{conservationlaw}) corresponds to the conservation law. For later convenience for calculations, we recast Eq.~(\ref{conservationlaw}) as 
\begin{eqnarray}
\sigma^\prime =-\frac{d-2}{a}(\sigma+p). \label{cons1}
\end{eqnarray}

We can get the conservation law of mechanical energy for the exotic matter on the thin shell throat by recasting Eq.~(\ref{energy density}) as follows:
\begin{eqnarray}
&&\dot {a}^2 + V(a) = 0, \label{energy cons} \\
&&V(a)=-\left(\frac{4\pi a \sigma}{d-2}\right)^2 - \left(\frac{f_+ -f_-}{2}\right)^2 \left(\frac{d-2}{8\pi a \sigma}\right)^2+\frac{1}{2}\left(f_++ f_-\right), \label{V(a)001}
\end{eqnarray}
where $V(a)$ is called the effective potential or just the potential.
From Eq.~(\ref{energy cons}), the range of $a$ which satisfies $V(a)\leq 0$ is the movable range for the shell. Since we obtained Eq.~(\ref{energy cons}) by twice squaring of Eq.~(\ref{energy density}), there is possibility that we take wrong solutions which satisfy Eq.~(\ref{energy cons}) but do not satisfy Eq.~(\ref{energy density}). See Appendix for the condition of right solutions. 
By differentiating Eq.~(\ref{energy cons}) with respect to $\tau$, we get the equation of motion for the shell as
\begin{eqnarray}
\ddot a=-\frac{1}{2}V^\prime (a). \label{eom of shell}
\end{eqnarray}

Suppose a thin shell throat be static at $a=a_0$ and its throat radius satisfy the relation	
\begin{eqnarray}
f(a_0)>0.  \label{f0>0}
\end{eqnarray}
This condition is called the horizon-avoidance condition in Ref~\cite{barcelo&visser}. 
We analyze stability against small perturbations preserving symmetries. To determine whether the shell is stable or not against the perturbation, we use Taylor expansion of the potential $V(a)$ around the static radius $a_0$ as
\begin{equation}
V(a)=V(a_0)+V^\prime(a_0)(a-a_0)+\frac{1}{2}V^{\prime \prime}(a_0)(a-a_0)^2+\mathcal O((a-a_0)^3) .\label{V(a)}
\end{equation}
From Eqs.~(\ref{energy cons}) and (\ref{eom of shell}), $\dot a_0=0,\ddot a_0=0\Leftrightarrow V(a_0)=0,V^\prime(a_0)=0$ at $a=a_0$ so the potential given by Eq.~(\ref{V(a)}) reduces to
\begin{equation}
V(a)=\frac{1}{2}V^{\prime \prime}(a_0)(a-a_0)^2+\mathcal O((a-a_0)^3). \label{V(a)2}
\end{equation}
Therefore, the stability condition against radial perturbations for the thin shell is given by
\begin{equation}
V^{\prime \prime}(a_0)>0. \label{V''>0}
\end{equation}

\section{Wormholes with a negative tension brane with Z2 symmetry} 
\label{Wormholes with a negative tension brane}
\subsection{Effective potential}
From now on, for simplicity, we assume Z2 symmetry, that is, we assume
$M_+=M_-$, $Q_+=Q_-$ and $\Lambda_+=\Lambda_-$ and hence $f_+(r)=f_-(r)$.
We denote $M_+=M_-=M$, $Q_+=Q_-=Q$, $\Lambda_+=\Lambda_-=\Lambda$ and 
$f(r):=f_+(r)=f_-(r)$. Then Eq.~(\ref{V(a)}) reduces to
\begin{eqnarray}
V(a)=f(a)-\left( \frac{4\pi a \sigma}{d-2}\right)^2. \label{V(a)01}
\end{eqnarray}
We investigate wormholes which consist of a negative tension brane. From Eqs.~(\ref{energy density}) and (\ref{surface pressure}), the surface energy density and surface pressure for the negative tension brane are represented as
\begin{eqnarray}
&& \sigma = -\frac {d - 2}{4\pi a} A=\alpha,  \label{energy density01} \\
&& p = \frac{1}{4\pi} \left( \frac{B}{A} + \frac{d - 3}{a}A\right)=-\alpha, \label{surface pressure01}
\end{eqnarray}
where $\alpha<0$. By substituting Eqs.~(\ref{energy density01}) and
(\ref{surface pressure01}) into Eq.~(\ref{cons1}), we obtain
$\sigma^\prime =0$ irrespective of $d$. Therefore, $\alpha$ must be
constant and then both $p$ and $\sigma$ must be constant, too. So the
effective potential reduces to
\begin{eqnarray}
V(a)=f(a)-\left(\frac{4\pi \alpha}{d-2}\right)^2a^2. \label{V(a)02}
\end{eqnarray}

\subsection{Static solutions, stability criterion and horizon-avoidance condition}
The present system may have static solutions $a=a_{0}$. We define $p_0:=p(a_0)$ and $\sigma_0:=\sigma(a_0)$, where $a_0$ satisfies Eq.~(\ref{f0>0}) :
\begin{eqnarray}
&& \sigma_0 = -\frac {d - 2}{4\pi a_0} A_0=\alpha,  \label{energy density02} \\
&& p_0 = \frac{1}{4\pi} \left( \frac{B_0}{A_0} + \frac{d - 3}{a_0}A_0\right)=-\alpha, \label{surface pressure02}
\end{eqnarray}
where
\begin{eqnarray}
A_0:=\sqrt{f(a_0)},B_0:=\frac{1}{2}f^\prime(a_0).
\end{eqnarray}
Eliminating $\alpha$ from Eqs.~(\ref{energy density02}) and (\ref{surface pressure02}), we can obtain the equation for the static solutions,
\begin{eqnarray}
\frac{a_0}{2}f^\prime(a_0)-f(a_0)=0. \label{static sol1}
\end{eqnarray}
The explicit form of Eq.~(\ref{static sol01}) is
\begin{eqnarray}
2ka_0^{2(d-3)} - (d-1)Ma_0^{d-3} + 2(d-2)Q^2=0. \label{static sol02}
\end{eqnarray}
The stability conditions for wormholes are shown in the previous section as $V^{\prime \prime}(a_0)>0$. The relation between $\alpha$ and $a_0$ is given by Eq.~(\ref{energy density02}). By substituting Eq.~(\ref{energy density02}) into Eq.~(\ref{V(a)02}), we obtain stability conditions as
\begin{eqnarray}
V^{\prime \prime}(a_0)&=& f^{\prime \prime}_0 -2\left( \frac{4\pi \alpha}{d-2} \right)^2 = f^{\prime \prime}_0-2\frac{f_0}{a_0^2}>0 \nonumber \\ 
&\Leftrightarrow& (d-3)\left[4k-(d-1)\frac{M}{a_0^{d-3}}\right]<0. \label{stable condition01}
\end{eqnarray}
We used Eq.~(\ref{static sol02}) to derive Eq.~(\ref{stable
condition01}). As one can see, since static solutions of
Eq.~(\ref{static sol02}) and stability conditions of Eq. (\ref{stable
condition01}) do not contain $\Lambda$, the cosmological constant only
affects the horizon-avoidance condition of Eq.~(\ref{f0>0}). By studying both the existence of static solutions and stability conditions, we can search static and stable wormholes. 

\subsection{$d=3$}
We first analyze static solutions and stability for $d=3$. 
In this case, the stability analysis is simple. The metric (\ref{f}) becomes
\begin{eqnarray}
f(a)=k-M+Q^2-\frac{\Lambda}{3}a^2,
\end{eqnarray}
so the potential is 
\begin{eqnarray}
V(a)=k-M+Q^2-\Lambda_{{\rm eff}}a^2,
\end{eqnarray}
where 
\begin{eqnarray}
\Lambda_{{\rm eff}}\equiv \frac{\Lambda}{3}+\left( \frac{4\pi \alpha}{d-2}\right)^2. \label{lambda eff01}
\end{eqnarray}
Since the shell is static, $V^\prime (a_0)=0$, we obtain 
\begin{eqnarray}
\Lambda_{{\rm eff}}=0. \label{lambda eff02}
\end{eqnarray}
Therefore the potential is $V(a)=k-M+Q^2$. Besides, $V(a_0)=0$ yields  $k-M+Q^2=0$ so we obtain
\begin{eqnarray}
f(a)=-\frac{\Lambda}{3}a^2, V(a)=0.
\end{eqnarray}
Therefore any radius $a_0$ is static. We find $V^{\prime
\prime}(a_0)=0$, which means the wormholes is marginally stable. The
horizon-avoidance condition $f(a_0)>0 \Leftrightarrow \Lambda<0$ is
satisfied because of (\ref{lambda eff01}) and (\ref{lambda eff02}). This
wormhole is constructed by pasting a couple of anti de-Sitter spacetimes.
 
\subsection{$d\ge 4$}
From now on, we assume $d\geq4$. For $k\neq0$ and $M\neq0$,
Eq.~(\ref{static sol02}) is a quadratic equation. The static solutions are then given by 
\begin{eqnarray}
a_{0\pm}^{d-3}= \frac{d-1}{4k}M( 1\pm b ), \label{k=+-1:static sol}
\end{eqnarray}
where
\begin{eqnarray}
b:= \sqrt{1-k\frac{q^2}{q_c^2}}, q:=\frac{|Q|}{|M|}, q_c:=\frac{(d-1)}{4\sqrt{d-2}}. \label{definition}
\end{eqnarray}
Combining Eqs.~(\ref{k=+-1:static sol}) and (\ref{stable condition01}),
we can see that for $b=0$, the positive and negative sign solutions 
coincide and their stability depends on higher order terms.
For $b>0$ and $k=+1$,
the negative sign solution is stable, while the positive
sign solution is unstable. 
For $k=-1$, we can conclude $b\ge 1$ and 
stability depends on the sign of mass $M$.
The horizon-avoidance condition (\ref{f0>0}) reduces to
\begin{eqnarray}
\frac{1}{3}\Lambda a_{0\pm}^2 < -\frac{(d-3)k}{(d-1)(d-2)(1\pm b)}\left[2-(d-1)(1\pm b)\right].  \label{f0>0 p2}
\end{eqnarray}
We investigate $k=+1$ and $k=-1$ cases, separately.

\subsubsection{$k=1$ and $M\ne 0$}
Though the original range for $b$ is $ 0\leq b \leq 1$, $b=0 \Leftrightarrow q=q_c$ does not satisfy the stability condition:
  For $q=q_c$, there is the only one static solution,
\begin{eqnarray}
a_0^{d-3}=\frac{d-1}{4}M. \label{m}
\end{eqnarray}
This double root solution is linearly marginally stable but 
nonlinearly unstable because $V'''(a_{0})\ne 0$.
From Eq.~(\ref{m}), we must have positive mass $M>0$ to make sure $a_0$ to be positive. The horizon-avoidance condition reduces to
\begin{eqnarray}
\lambda<R(d),
\end{eqnarray}
where $\lambda$ is a dimensionless quantity corresponding to $\Lambda$ and $R(d)$ is defined by
\begin{eqnarray}
\lambda:= \frac{\Lambda}{3}|M|^{\frac{2}{d-3}}, R(d)\equiv \left( \frac{4}{d-1} \right)^{\frac{2}{d-3}} \frac{(d-3)^2}{(d-1)(d-2)}. \label{R}
\end{eqnarray}
Since $R(d)$ is positive, we can have the wormhole even without $\Lambda$. 

If $b=1 \Leftrightarrow Q=0$, the negative sign solution vanishes. The positive sign solution which does not satisfy the stability condition is
\begin{eqnarray}
a_{0+}^{d-3}=\frac{d-1}{2}M. \label{m1}
\end{eqnarray}
From Eq.~(\ref{m1}), $M$ must be positive. One can verify that the horizon-avoidance condition cannot be satisfied, $f(a_0)<0$.

\begin{figure}[htbp]
  \begin{center}
    \includegraphics[bb=0 0 530 327,width=10.5cm]{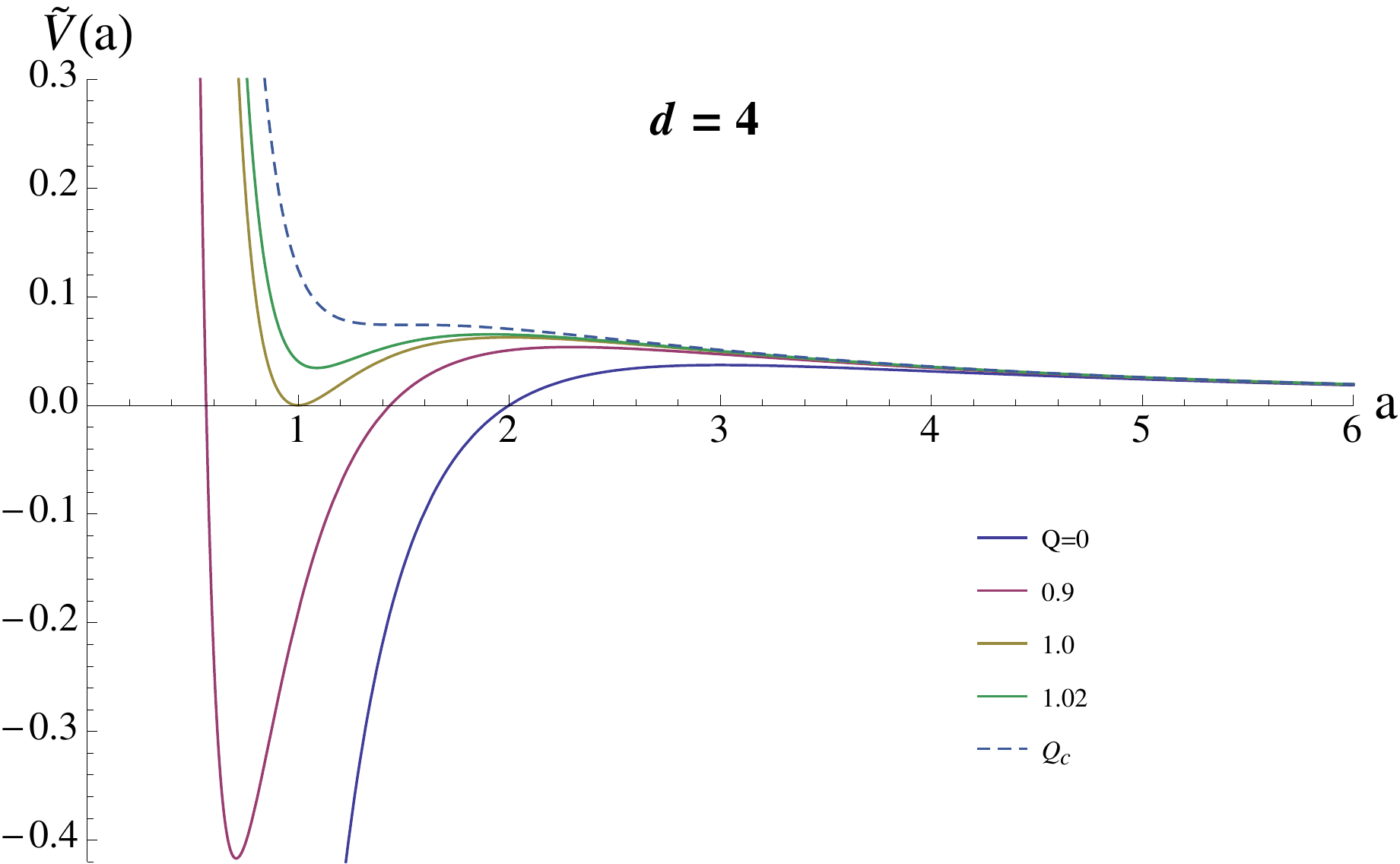}
    \caption{The potential $\tilde V (a)$ for $d=4$, $k=+1$ and $M=2$. The dashed line is the potential for the critical value defined in Eq.~(\ref{definition}).}
    \label{d=4}
  \end{center}
\end{figure}
\begin{figure}[htbp]
  \begin{center}
    \includegraphics[bb=0 0 530 327,width=10cm]{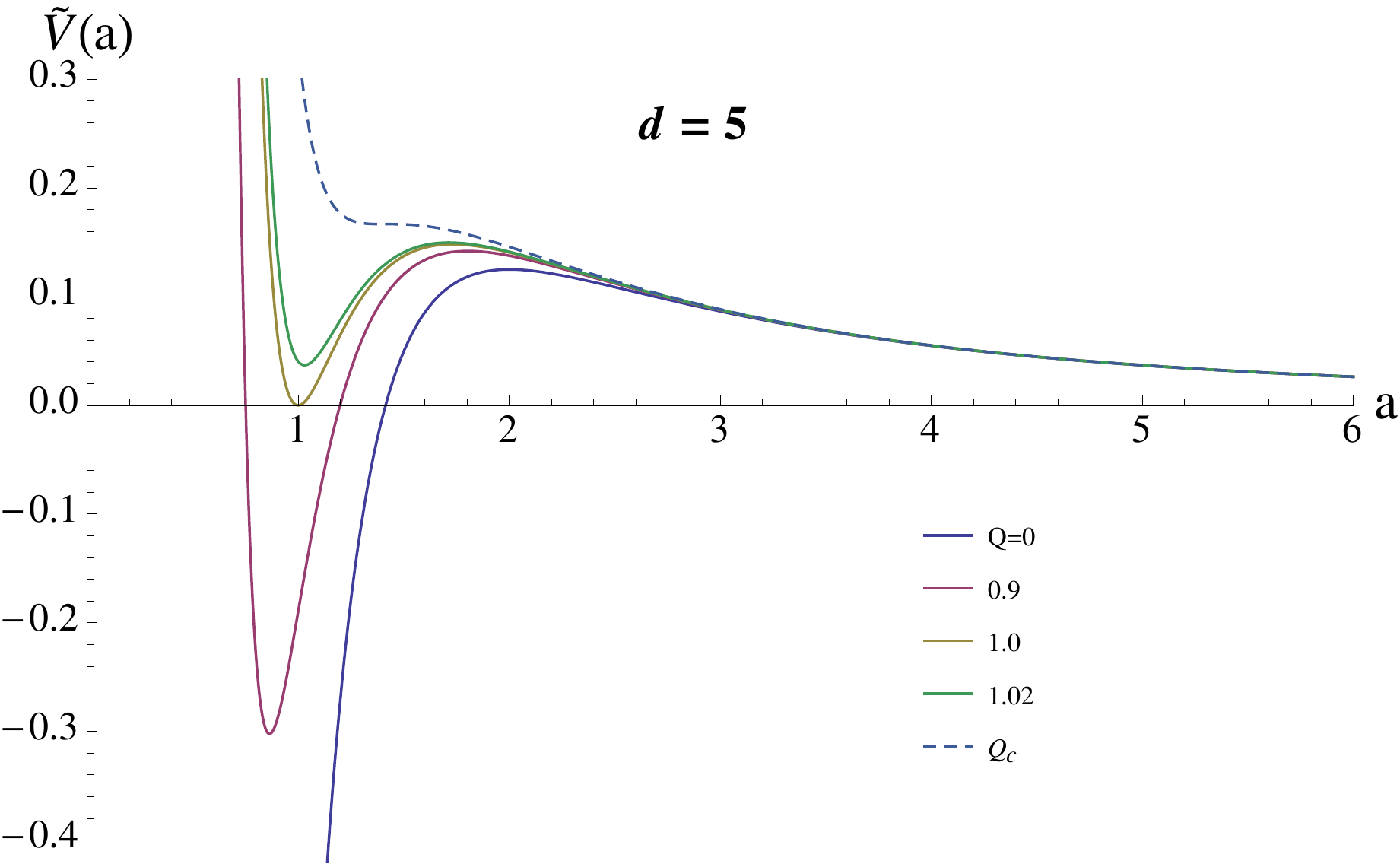}
    \caption{The potential for $d=5$, $k=+1$ and $M=2$.}
    \label{d=5}
  \end{center}
\end{figure}
\begin{figure}[tbp]
  \begin{center}
    \includegraphics[bb=0 0 670 409,width=11cm]{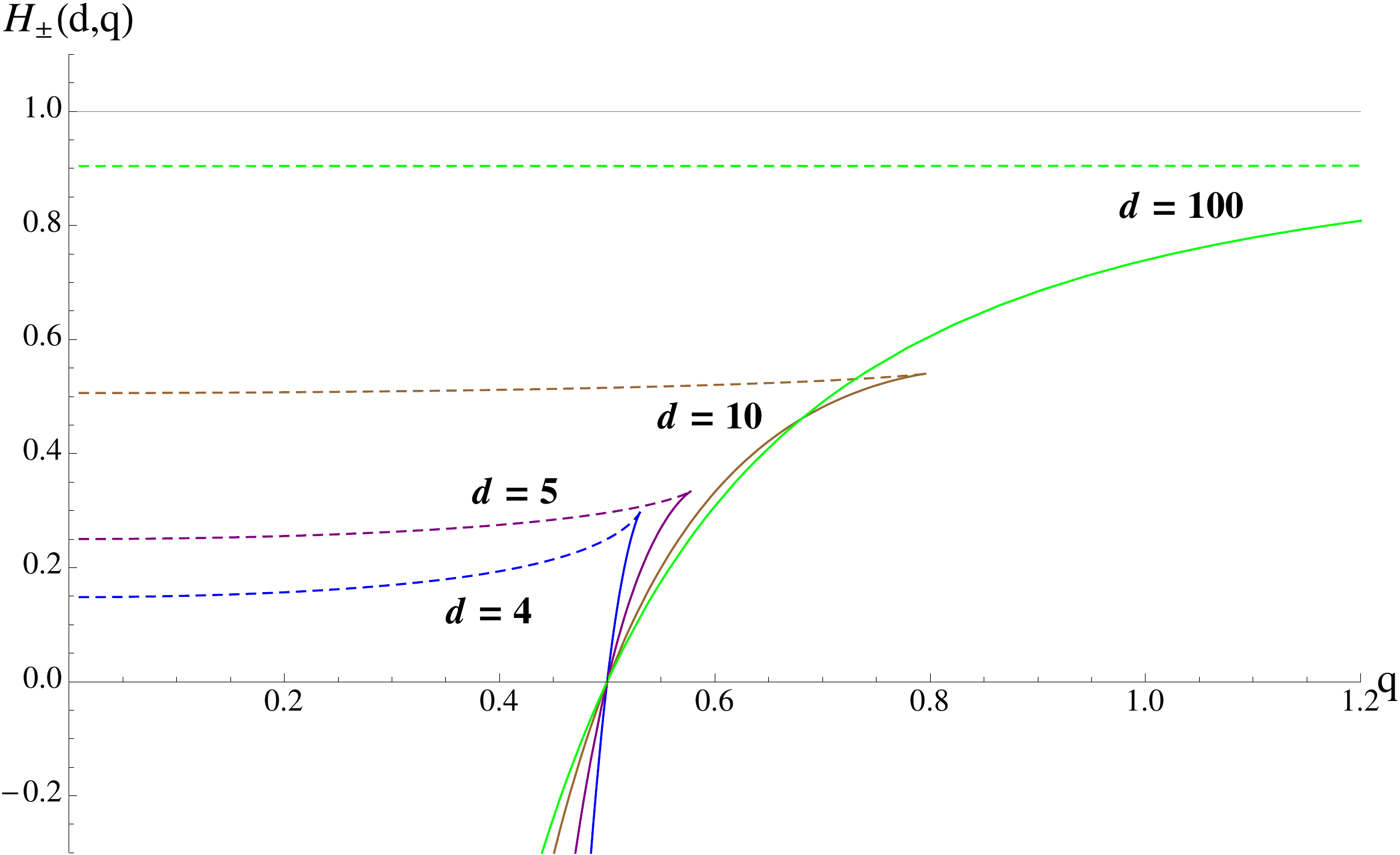}
    \caption{
The functions $H_{\pm}(d,q)$ defined in Eq.~(\ref{ABC}) are plotted.
The dashed lines are $H_+(d,q)$ and the solid lines are $H_-(d,q)$. The
   regions below these curves represent the regions of $\lambda$ which satisfy Eq.~(\ref{H}). When $q<1/2$, $H_+(d,q)$ is positive, while $H_-(d,q)$ is negative. When $1/2<q<q_c$,  both $H_\pm(d,q)$ are positive. As the number of dimensions and charge increase, it approaches $1$.}
\label{fig:H+-}
  \end{center}
\end{figure}

When $0<b<1\ \Leftrightarrow  0 < q<q_c$, there are two static solutions
\begin{eqnarray}
a_{0\pm}^{d-3}= \frac{d-1}{4}M(1\pm b). \label{static sol01}
\end{eqnarray}
The stability condition is satisfied if we take the negative sign
solution of Eq.~(\ref{static sol01}). The positive sign solution is
unstable. Since the static solution must be positive, we must have
$M>0$. The following transformation helps us to understand the potentials:
\[
\tilde V (a)\equiv
\frac{k}{a^2}-\frac{M}{a^{d-1}}+\frac{Q^2}{a^{2(d-2)}} 
\]
so that 
\[
\dot
a^2+V(a)=0 \Leftrightarrow \left( \frac{\ln a}{d\tau} \right)^2+\tilde V
(a)=\Lambda_{\rm eff}.
\]
The potentials $\tilde{V}(a)$ are plotted in Figs.~\ref{d=4} and ~\ref{d=5}
for $d=4$ and $d=5$, respectively. 
The horizon-avoidance condition (\ref{f0>0 p2}) reduces to
 \begin{eqnarray}
\lambda  < H_\pm(d,q), \label{H}
\end{eqnarray}
where
\begin{eqnarray}
H_\pm(d,q) \equiv-\left\{ \frac{4}{(d-1)(1\pm b)} \right\}^{\frac{2}{d-3}}\frac{d-3}{(d-1)(d-2)(1\pm b)}\left[2-(d-1)(1\pm b)\right]. \label{ABC}
\end{eqnarray}
The positive and the negative sign corresponds with the sign of Eq.~(\ref{static sol01}) in same order.
When one take the positive sign, $H_+(d,q)$, one can find the inside of the square brackets of Eq.~(\ref{ABC}) is negative, then $H_+(d,q)$ is positive.  Therefore Eq.~(\ref{H}) is satisfied even with $\Lambda=0$ in the case of $H_+(d,q)$. 
Similarly, in the case of the negative sign, $H_-(d,q)$, if the inside of the square brackets of Eq.~(\ref{ABC}) can be negative, Eq.~(\ref{H}) is satisfied even with $\Lambda=0$. In this stable case, one can achieve this situation if and only if
\begin{eqnarray}
\frac{1}{2}<q<q_c \label{ABCD}
\end{eqnarray}
is satisfied. $H_{\pm}(d,q)$ are plotted in Fig.~\ref{fig:H+-}. 
Therefore, we can construct a stable TSW without $\Lambda$ when the condition Eq.~(\ref{ABCD}) is satisfied. So the extremal or subextremal charge $q \leq1/2$ of the Reissner-Nordstr\"om spacetime cannot satisfy Eq.~(\ref{ABCD}). We can reconfirm the previous result by taking $d=4$ and $M=2m$ for Eq.~(\ref{ABCD}) as
\begin{eqnarray}
1<\left(\frac{|Q|}{m}\right)^2 < \frac{9}{8}.
\end{eqnarray}
This coincides with the previous result by Barcel\'o and
Visser~\cite{barcelo&visser}. From Eq.~(\ref{ABCD}), as $d$ increases,
larger charge is allowed to construct a stable wormhole without
$\Lambda$. This class of wormholes are constructed by pasting a couple
of over-charged higher dimensional Reissner-Nordstr\"om spacetimes.

\subsubsection{$k=-1$ and $M\ne 0$}
\begin{figure}[htbp]
  \begin{center}
    \includegraphics[bb=0 0 452 274,width=10cm]{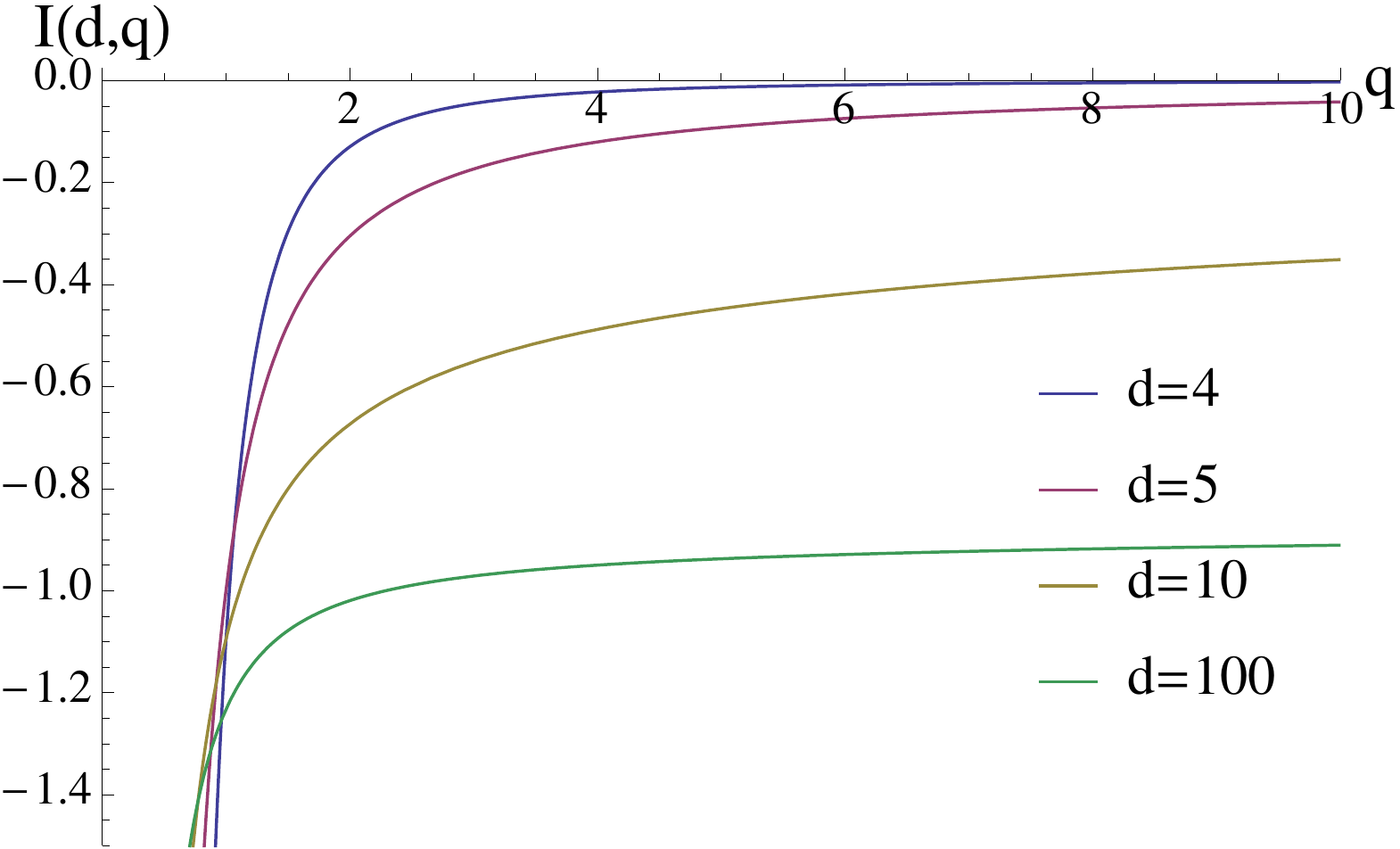}
    \caption{
The function $I(d,q)$ defined in Eq.~(\ref{I}) is plotted.
The regions below these
   curves represent the regions of $\lambda$ which satisfy
   Eq.~(\ref{avoidance01}). $I(d,q)$ is negative. In $d=4$, $I(4,q)\to0$
   as $q\to \infty$.
\label{fig:lambda}}
  \end{center}
\end{figure}
The static solution is given by 
\begin{eqnarray}
a_{0\pm}^{d-3}=- \frac{d-1}{4}M( 1\pm b ), \label{k=-1:static sol02}
\end{eqnarray}
where
\begin{eqnarray}
b= \sqrt{1+\frac{q^2}{q_c^2}}.
\end{eqnarray}
$b$ is more than than or equal to unity i.e. $b\geq 1$.  
If $M>0$ and $Q\neq0$, the stability condition Eq.~(\ref{stable
condition01}) is satisfied, if we take the negative sign solution of
Eq.~(\ref{k=-1:static sol02}). If $M>0$ and $Q=0$, since the negative
sign solution vanishes and the positive sign solution is negative, there
is no static solution. Therefore, if $M>0$ and $Q\neq0$, we can have 
stable wormholes. In this case the horizon-avoidance condition Eq.~(\ref{f0>0 p2}) reduces to
\begin{eqnarray}
\lambda <I(d,q), \label{avoidance01}
\end{eqnarray}
where 
\begin{eqnarray}
I(d,q):= -\frac{d-3}{(d-1)(d-2)}\left\{ \frac{4}{(d-1)(b-1)} \right\}^{-\frac{2}{d-3}} \left[\frac{2}{b-1} + d-1 \right]. \label{I}
\end{eqnarray}
Eq.~(\ref{avoidance01}) can be satisfied only if $\Lambda$ is negative and
$|\Lambda|$ is sufficiently large. One can find that the inside of the
square brackets on the right hand side of Eq.~(\ref{I}) is positive, so
the vanishing cosmological constant cannot satisfy Eq.~(\ref{I}) 
unlike for $k=+1$. 
The value of $I(d,q)$ determines the maximum value for $\lambda$ which
is needed to achieve the horizon-avoidance condition
Eq.~(\ref{avoidance01}). $I(d,q)$ is plotted in Fig.~\ref{fig:lambda}.
Since $I(d,q)<0$, we need a negative cosmological constant 
for the horizon-avoidance condition to be satisfied.

\begin{figure}[tbp]
  \begin{center}
    \includegraphics[bb=0 0 595 371,width=9cm]{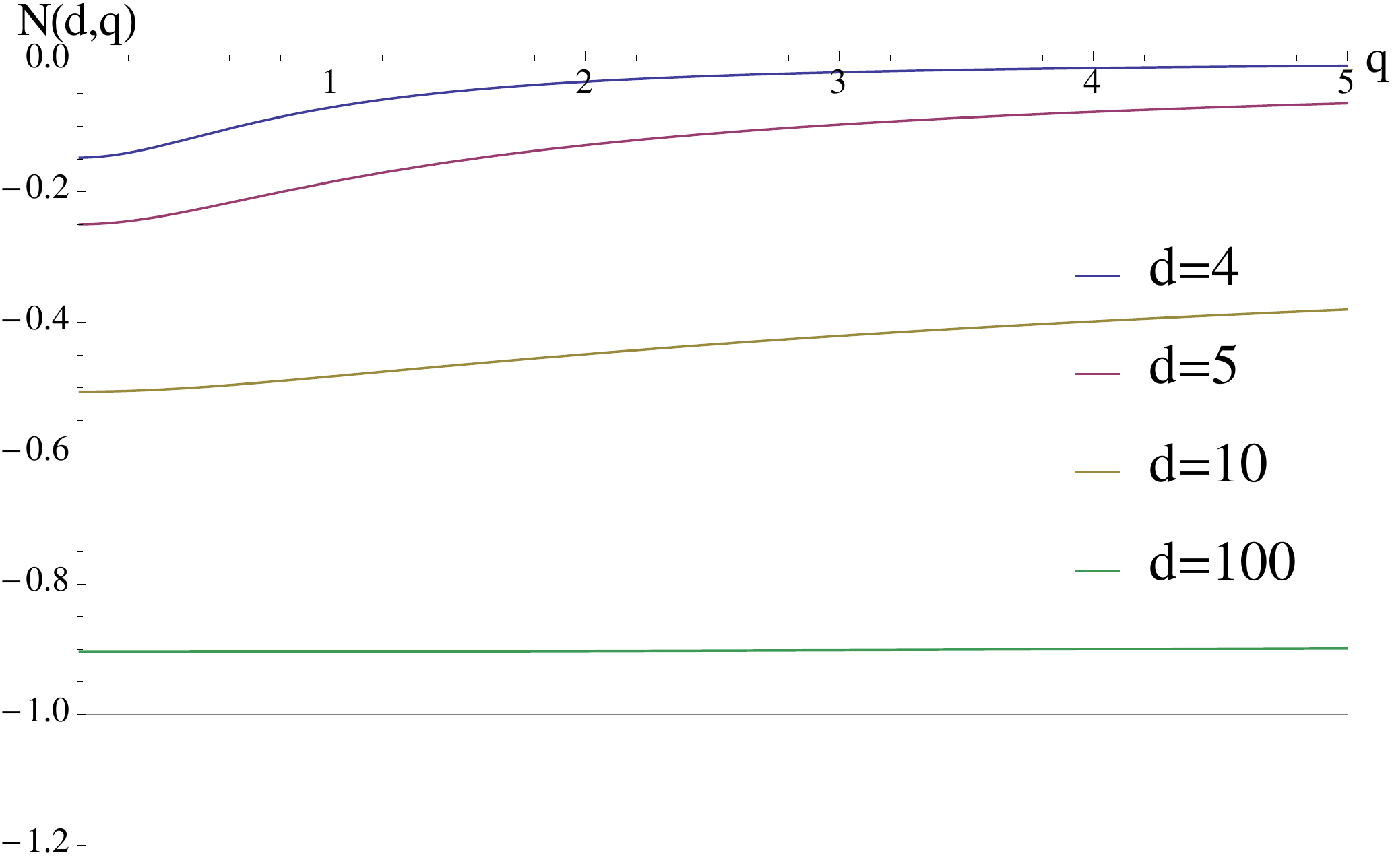}
    \caption{
The function $N(d,q)$ defined in Eq.~(\ref{N}) is plotted.
The regions below these curves represent the regions of $\lambda$ which satisfy Eq.~(\ref{avoidance02}). $N(d,q)$ is negative. As the number of dimensions increases, it approaches $-1$.}
    \label{fig:N}
  \end{center}
\end{figure}

If $M<0$, the stability condition is satisfied if we take the positive
sign solution Eq.~(\ref{k=-1:static sol02}) whether it is 
with or without charge. 
The negative sign solution contradicts $a_{0}>0$.
The positive sign solution is 
\begin{eqnarray}
a_{0+}^{d-3}= \frac{d-1}{4}|M|( 1+ b ). \label{Nstatic}
\end{eqnarray}
In this case, the horizon-avoidance condition Eq.~(\ref{f0>0 p2}) reduces to
\begin{eqnarray}
\lambda<N(d,q), \label{avoidance02}
\end{eqnarray}
where 
\begin{eqnarray}
N(d,q) \equiv \left\{ \frac{4}{(d-1)(1+ b)} \right\}^{\frac{2}{d-3}}\frac{d-3}{(d-1)(d-2)(1+ b)}\left[2-(d-1)(1+ b)\right].  \label{N}
\end{eqnarray}
$N(d,q)$ is plotted in Fig.~\ref{fig:N}. Since $N(d,q)<0$, we find that a negative cosmological constant is needed to achieve the horizon avoidance. 

\subsubsection{$k\neq0$ and $M=0$}
For $k=+1$, from Eq.~(\ref{static sol02}), we find there is no static solution.
For $k=-1$, 
Eq.~(\ref{static sol02}) has a double root solution: 
\begin{eqnarray}
a_0^{d-3}=\sqrt{d-2}|Q|, \label{k=-1,M=0}
\end{eqnarray}
where $Q\neq0$ must hold for the positivity of $a_0$. One can easily verify the stability condition is satisfied in this case. The horizon avoidance Eq.~(\ref{f0>0}) reduces to 
\begin{eqnarray}
\frac{\Lambda}{3}<S(d,q), \label{AA01}
\end{eqnarray}
where
\begin{eqnarray}
S(d,q)\equiv -|Q|^{-\frac{2}{d-3}}(d-3)(d-2)^{d-4} \label{S}
\end{eqnarray}
Since the right hand side of Eq.~(\ref{AA01}) is negative, we need a
negative cosmological constant for the stable wormhole in this
case. However, even an arbitrarily small $|\Lambda|$ can 
satisfy Eq.~(\ref{AA01}), if $|Q|$ is sufficiently large. 

\subsubsection{$k=0$ and $M\neq0$}

\begin{figure}[tbp]
  \begin{center}
    \includegraphics[bb=0 0 439 270,width=9cm]{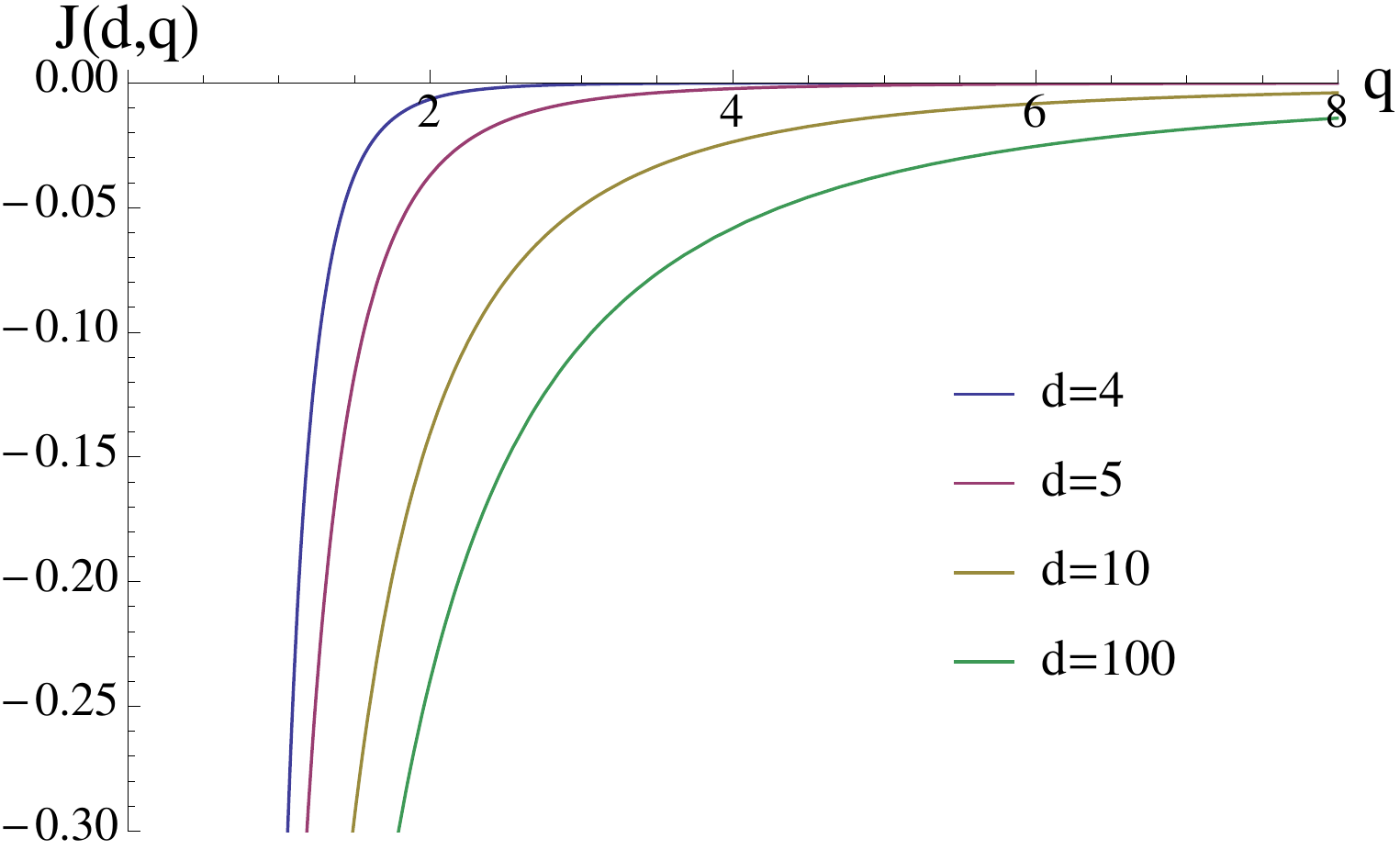}
    \caption{The function $J(d,q)$ defined in Eq.~(\ref{J}) is plotted. 
The regions below
   these curves represent the regions of $\lambda$ which satisfy
   Eq.~(\ref{k=0:out of horizon}). $J(d,q)$ is negative. As the number
   of dimensions increases, it approaches $-1/4q^2$.}
    \label{fig:J}
  \end{center}
\end{figure}

There is the only one static solution that is
\begin{eqnarray}
a_0^{d-3}= 2\frac{d-2}{d-1}\frac{|Q|^2}{M}. \label{k=0:static sol}
\end{eqnarray}
$M>0$ and $Q\neq0$ must hold since $a_0$ must be positive. The stability condition is satisfied in this case. The horizon-avoidance condition reduces to
\begin{eqnarray}
\lambda <J(d,q), \label{k=0:out of horizon}
\end{eqnarray}
where
\begin{eqnarray}
J(d,q)\equiv -\frac{1}{4}\left(\frac{1}{q}\right)^{2\frac{d-1}{d-3}}\left( \frac{d-1}{2(d-2)} \right)^{\frac{2}{d-3}} \frac{(d-1)(d-3)}{(d-2)^2}. \label{J}
\end{eqnarray}
Since the right hand side of (\ref{k=0:out of horizon}) is negative, $\Lambda$ should be negative. Taking a limit of $d \to \infty$ leads to
\begin{eqnarray}
\lambda \left( \frac{|Q|}{M}\right)^2<-\frac{1}{4} \ \ {\rm as}\ \ d \to \infty. \label{k=0:limit of out of horizon}
\end{eqnarray}
The function $J(d,q)$ is plotted in Fig.~\ref{fig:J}.
Since $J(d,q)$ is negative, we need a negative cosmological
constant for the horizon avoidance.

\subsubsection{$k=0$ and $M=0$}
In this case, we must have $Q=0$ to satisfy Eq.~(\ref{static sol02}). Then we get $V(a)=-\Lambda_{{\rm eff}}a^2$, where $\Lambda_{{\rm eff}}$ is defined in Eq.~(\ref{lambda eff01}). Since the shell is static, $V^\prime (a_0)=0$, we find
\begin{eqnarray}
\Lambda_{{\rm eff}}=0 \Leftrightarrow \frac{\Lambda}{3}=-\left( \frac{4\pi \alpha}{d-2}\right)^2.  \label{lamuda}
\end{eqnarray}
Then, the potential vanishes, i.e., $V(a)=0$, which means the wormhole
can be static at an arbitrarily radius and is marginally stable. Since the cosmological constant turned out to be negative from Eq.~(\ref{lamuda}), 
we have $f(a)=(4\pi \alpha \slash d-2)^2a^2$, then the horizon avoidance
$f(a_0)>0$ is automatically satisfied. This solution is what we can call
an another side of 
Randall-Sundrum (RS) II brane world model~\cite{randall-sundrum}. In both
cases, the ingredients are a couple of anti de-Sitter (AdS) spacetimes. The 
RS II model is a compactified spacetime by pasting the {\it interiors}
of a couple of AdS spacetimes at the boundaries, while the
wormhole solution is a non-compactified spacetime by pasting 
the {\it exteriors} of a couple of AdS spacetimes at the boundaries.

\section{Summary and Discussion}

We developed the thin shell formalism for $d$ dimensional spacetimes
which is more general than Dias and Lemos
formalism~\cite{Dias&Lemos}. We investigated spherically, planar
(cylindrically) and hyperbolically symmetric wormholes with a pure 
negative tension brane and found and classifies Z2 symmetric 
static solutions which are stable
against radial perturbations. We found that in most cases
charge is needed to keep the static throat radius positive and that a
negative cosmological constant tends to decrease the radius of the black
hole horizon and then to achieve the horizon avoidance. So the
combination of an electric
charge and a negative cosmological constant makes it easier to construct
stable wormholes. However, a negative cosmological constant is
unnecessary in a certain situation of $k=+1$ and $M>0$ and charge is
unnecessary in a certain situation of $k=-1$ and $M<0$. 
We summarize the results in Tables \ref{table d=3}, \ref{table k=+1}, \ref{table k=-1} and \ref{table k=0}.

In three dimensions, there is only possibility to have a marginally
stable wormhole. The ingredients of this wormhole are a couple of 
AdS space-times.

Then, we restrict the spacetime dimensions to be higher than or equal to
four. For $k=+1$, spherically symmetric thin shell wormholes which are
made with a negative tension brane are investigated. It turns out that
the mass must be positive, i.e., $M>0$. The obtained wormholes can be
interpreted as the higher dimensional counterpart of Barcel\'o-Visser
wormholes~\cite{barcelo&visser}. As a special case, if
$1/2<q<q_{c}$, one can obtain a stable wormhole without a
cosmological constant. This wormhole consists of a negative tension 
brane and a couple of 
over-charged Reissner-Nordstr\"om space-times.

For $k=-1$, though it is hard to imagine how such symmetry is physically
realized, they are interesting from the
viewpoint of stability analyses. It turns out that $M$ can be positive, zero and negative for stable wormholes.
In this geometry, there is no upper limit for $|Q|$ for 
stable wormholes. There is possibility for a stable wormhole without charge if $M<0$ and $\lambda<N(d,0)$ is satisfied.

For $k=0$, the geometry is planar symmetric or cylindrically symmetric. 
In this {\bf case}, since the generalized Birkhoff's theorem
does not apply~\cite{sato-kodama}, we should regard 
the Reissner-Nordstr\"om-(anti) de Sitter spacetime as a special
solution to the electrovacuum Einstein equations. 
This means that the present analysis only covers a part of possible 
static thin shell wormholes and the stability against only a part 
of possible radial perturbations.
Under such a restriction, we find that 
we need $Q\neq0$ and $\Lambda<0$ to have stable wormholes. There is no upper limit for $|Q|$. In the zero mass case, the wormhole is marginally stable.

Finally, we would note that the existence and stability of
negative tension branes as thin shell wormholes crucially on the curvature of 
the maximally symmetric $(d-2)$ dimensional manifolds. 
On the other hand, they do not qualitatively
but only quantitatively depend on the number of spacetime dimensions.

\acknowledgements
The authors would like to thank Tsutomu Kobayashi, Hideki Maeda,
Ken-Ichi Nakao, Shinya Tomizawa, Takashi Torii and Chul-Moon Yoo for helpful comments and discussions.  
T.K. was supported in part by Rikkyo University Special Fund for Research.
T.H. was partially supported by the Grant-in-Aid No. 26400282 for Scientific
Research Fund of the Ministry of Education, Culture, Sports, Science and Technology, Japan.

\appendix
\section{The condition for right solutions}
We obtained the equation of motion (\ref{energy cons}) by twice squaring Eq. (\ref{energy density}).
Eq.~(\ref{energy density}) is equivalent to 
\begin{eqnarray}
  \begin{cases}
    C^2\sigma^2=(A_++A_-)^2,& \\
    \sigma<0.&
  \end{cases} \label{app sigma02}
\end{eqnarray}
The above is also equivalent to 
\begin{eqnarray}
 \begin{cases}
	\left(C^2\sigma^2-A_+^2-A_-^2\right)^2 =(2A_+A_-)^2, & \\
	C^2\sigma^2-A_+^2-A_-^2>0, & \\ 
	\sigma<0. &
 \end{cases} \label{app sigma03} 
\end{eqnarray}
The first equation of Eqs.~(\ref{app sigma03}) is obtained by squaring
the first equation of Eqs.~(\ref{app sigma02}). By recasting the first
equation of Eqs.~(\ref{app sigma03}), we obtain the equation of motion
Eq. (\ref{energy cons}). However, the solution of Eq. (\ref{energy cons})
is valid if and only if the second and third inequalities of Eqs.~(\ref{app sigma03}) are satisfied. The second inequalities of Eqs. (\ref{app sigma03}) is represented explicitly as
\begin{eqnarray}
A_{0}^{2}>0,
\end{eqnarray}
where Eq.~(\ref{energy density02}) is used.
Since $A_{0}^{2}=f(a_{0})>0$ is guaranteed by the horizon-avoidance
condition, our analysis does not contain wrong solutions 
of Eq.~(\ref{energy cons}).
On the other hand, if we deal with a solution violating the
horizon-avoidance condition, we need to reconsider whether it is
valid in more general context.

\newpage

\begin{table}[htb]
\begin{center}
\caption{The existence and stability of Z2 symmetric} 
static wormholes in three
 dimensions. $k=1$, $0$ and $-1$ correspond to spherical, planar
 (cylindrical) and hyperbolic symmetries, respectively.
\label{table d=3}
\begin{tabular}{|c|c|c|c|}\hline
 & Static solution & Horizon avoidance & Stability \\ \hline 
$k-M+Q^2=0$ & $\forall a_0>0$ & Satisfied & Marginally stable \\ \hline
$k-M+Q^2 \neq0$ & None & -- & -- \\ \hline
\end{tabular}
\end{center} 
\end{table}

\begin{table}[htb]
\begin{center}
\caption{The existence and stability of Z2 symmetric 
static wormholes in spherical
 symmetry in four and higher dimensions. $q$, $\lambda$ and $q_{c}$ are
 defined as $q:=|Q/M|$, $\lambda:= (\Lambda/3)|M|^{\frac{2}{d-3}}$ and 
 $q_c:=(d-1)/(4\sqrt{d-2})$, respectively.
The expressions for the static solutions $a_{0\pm}$ ($0<a_{0-}<a_{0+}$) are 
given by Eqs.~(\ref{k=+-1:static sol}) and (\ref{definition}) with
 $k=1$. 
$H_\pm(d,q)$ are given by Eq.~(\ref{ABC}) and plotted in
 Fig.~\ref{fig:H+-}. $R(d)$ is
 positive and given by Eq.~(\ref{R}).
Note that $H_{+}>0$ for
 $0\le q\le q_{c}$, while $H_{-}>0$ only for $1/2<q \le q_{c}$. 
Therefore, if $\Lambda=0$, the horizon-avoidance condition holds 
for $a_{0+}$ for $0<q\le q_{c}$, while it does for $a_{0-}$ only 
for $1/2<q\le q_{c}$. For $M>0$ and $q=q_{c}$, the double root solution
 $a=a_{0\pm}$ is linearly marginally stable but nonlinearly unstable.
\label{table k=+1}
}
\begin{tabular}{|c|c|c|c|c|}\hline
\multicolumn{2}{|c|}{} & {Static solution} & Horizon avoidance & Stability \\ \hline\hline
& $q=0$ & $[(d-1)M/2]^{1/(d-3)}$ & Not satisfied & Unstable \\ \cline{2-5}
$M>0$ & $0<q<q_c$ &
	 $a_{0\pm}$ &
     $\lambda<H_{\pm}(d,q)$ for $a_{0\pm}$&
     $a_{0-}$: Stable, $a_{0+}$: Unstable \\
\cline{2-5} & $q=q_c$ & $[(d-1)M/4]^{1/(d-3)}$  & $\lambda<R(d)$ & 
	     Unstable \\ \cline{2-5}
& $q_{c}<q$ & None & -- & --  \\ \hline
\multicolumn{2}{|c|}{$M<0$}  & None & -- & --   \\ \hline
\multicolumn{2}{|c|}{$M=0$} & None & -- & --  \\ \hline
\end{tabular}
\end{center} 
\end{table}
\begin{table}[htbp]
\begin{center}
\caption{The existence and stability of Z2 symmetric static wormholes in
 hyperbolic symmetry in four and higher dimensions. The definitions
for $q$ and $\lambda$ are same as in Table~\ref{table k=+1}.
The expressions for the static solutions $a_{0\pm}$ 
are given by Eqs.~(\ref{k=+-1:static sol}) and (\ref{definition}) with
 $k=-1$. 
$I(d,q)$ and $N(d,q)$ are given by 
Eqs.~(\ref{I}) and (\ref{N}) and plotted in
 Figs.~\ref{fig:lambda} and \ref{fig:N}, respectively.
$S(d,q)$ is given by Eq.~(\ref{S}).
 Since all of $I$, $N$ and $S$ are negative, the 
horizon-avoidance condition cannot be satisfied with $\Lambda=0$
for any cases in hyperbolic symmetry.
\label{table k=-1}}
\begin{tabular}{|c|c|c|c|c|}\hline
 \multicolumn{2}{|c|}{}&{Static solution} & Horizon avoidance& Stability 
 \\ \hline\hline 
$M>0$ & $q=0$ & None & -- & -- \\ \cline{2-5}
& $q>0$ & $a_{0-}$ & $\lambda<I(d,q)$& Stable \\ \hline
$M<0$  & $q=0$ & $[(d-1)|M|/2]^{1/(d-3)}$  & $\lambda <N(d,0)$ & Stable\\ \cline{2-5}
&$q>0$ & $a_{0+}$ & $\lambda<N(d,q)$& Stable \\ \hline
$M=0$  & $Q=0$ & None & -- & -- \\ \cline{2-5}
& $|Q|>0$ & $[\sqrt{d-2}|Q|]^{1/(d-3)}$ & $\Lambda/3<S(d,q)$& Stable \\ \hline
\end{tabular}
\end{center} 
\end{table}
\begin{table}[htbp]
\begin{center}
\caption{The existence and stability of Z2 symmetric 
static wormholes in planar or
 cylindrical symmetry in four and higher dimensions. The definitions
for $q$ and $\lambda$ are same as in Table~\ref{table k=+1}.
Note that we assume that the bulk spacetime is described by the 
Reissner-Nordst\"om-(anti) de Sitter metric or its higher dimensional 
counterpart. $J(d,q)$ is given by Eq.~(\ref{J}) and plotted in
 Fig.~\ref{fig:J}.
Since $J$ is negative, the horizon-avoidance condition cannot be 
satisfied with $\Lambda=0$ 
for $M>0$ and $q>0$ in planar or cylindrical symmetry.
\label{table k=0}}
\begin{tabular}{|c|c|c|c|c|}\hline
\multicolumn{2}{|c|}{}&{Static solution} & Horizon avoidance& Stability 
 \\ \hline\hline 
$M>0$& $q=0$ & None & -- & -- \\ \cline{2-5}
& $q>0$ &$[2(d-2)q^{2}M/(d-1)]^{1/(d-3)}$ & $\lambda<J(d,q)$& Stable \\ \hline
$M<0$  & $q=0$ & None & -- & --\\ \cline{2-5}
&$q>0$ & None & -- & --\\ \hline
$M=0$ & $Q=0$ &$\forall a_0>0$ & Satisfied & Marginally stable \\ \cline{2-5}
& $|Q|>0$ & None & -- & --\\ \hline
\end{tabular}
\end{center} 
\end{table}


\begin{thebibliography}{99}
\bibitem{morris&thorne}M.~S.~Morris and K.~S.~Thorne, Am. J. Phys. {\bf
	56}, 395 (1988).
\bibitem{visser} M.~Visser, {\it Lorentzian Wormholes}, AIP Press, New York(1996);\\
M.~Visser, Phys.~ Rev. D {\bf 39}, 3182 (1989);\\
M.~Visser, Nucl. Phys. B {\bf 328}, 203 (1989).
\bibitem{poisson&visser} E.~Poisson and M.~Visser, Phys. Rev. D {\bf52}, 7318(1995).
\bibitem{Reissner} E.~F.~Eiroa and G.~E.~Romero, Gen
	Relativ. Gravit. {\bf 36}, 651 (2004);\\
E.~F.~Eiroa, Phys. Rev. D {\bf 78}, 024018 (2008).
\bibitem{Ishak&Lake} M.~Ishak and K.~Lake, Phys. Rev. D {\bf 65}, 044011
	(2002).
\bibitem{cosmologicalconstant}F.~S.~N.~Lobo and P.~Crawford, Class. Quantum Grav. {\bf21}, (2004) 391-404
\bibitem{cylindricalTSW} E.~F.~Eiroa and C.~Simone, Phys. Rev. D {\bf70}, 044008 (2004);\\
C.~Bejarano, E.~F.~Eiroa and C.~Simone, Phys. Rev. D {\bf75}, 027501 (2007);\\
E.~F.~Eiroa and C.~Simone, Phys. Rev. D {\bf81}, 084022 (2010);\\
E.~F.~Eiroa and C.~Simone, Phys. Rev. D {\bf82}, 084039 (2010);\\
M.~G.~ Richarte, Phys. Rev. D {\bf87}, 067503 (2013);\\
S.~H.~Mazharimousavi, M.~Halilsoy and Z.~Amirabi, Phys.~ Rev.~ D {\bf89}, 084003 (2014).
\bibitem{Garcia+} M.~N.~Garcia, F.~S.~N.~Lobo and M.~Visser, Phys.~ Rev.~ D {\bf86}, 044026(2012).
\bibitem{Dias&Lemos} G.~A.~S.~Dias and J.~P.~S.~Lemos, Phys.~ Rev.~ D {\bf 82},084023(2010).
\bibitem{scalarfield} H.~G.~Ellis, J.~ Math.~ Phys.~ {\bf 14}, 104 (1973);\\
K. A. Bronnikov, Acta Phys. Polon. B {\bf4},251 (1973).
\bibitem{Ellis instability} H.~Shinkai and S.~A.~Hayward, Phys.~ Rev.~ D {\bf66}, 044005 (2002);\\
J.~A.~Gonz\'alez, F.~S.~Guzman and O.~Sarbach, Class.~ Quantum Grav.~ {\bf26}, 015010 (2009);\\
J.~A.~Gonz\'alez, F.~S.~Guzman and O.~Sarbach, Class.~ Quantum Grav.~ {\bf26}, 015011 (2009).
\bibitem{barcelo&visser} C.~Barcel\'o and M.~Visser, Nucl.~ Phys.~ B {\bf584}, 415-435 (2000).
\bibitem{toolkit} E.~Poisson, \textit{A Relativist's Toolkit}, Cambridge University Press (2007) chap 3.
\bibitem{sato-kodama} V. Ruban, Gen. Relat. Grav. GR 8 Waterloo, 303 (1977);\\
K. Bronnikov and M. Kovalchuk, J. Phys. A: Math. Gen. {\bf13}, 187 (1980);\\
K. Bronnikov and V. Melnikov, Gen. Relat. Grav. {\bf27}, 465 (1995);\\
H. Goenner, Commun. Math. Phys. {\bf16}, 34-47 (1970);\\
H.~Sato and H.~Kodama, \textit{Ippan sotaisei riron}, Iwanami Shoten (1992) Chap.~5.
\bibitem{gravitation} C.~Misner, K.~Thorne and J.~Wheeler,
	\textit{GRAVITATION}, W.~H.~Freeman and company (1973) Part IV p.~554.
\bibitem{randall-sundrum} L.~Randall and R.~Sundrum, Phys.~ Rev.~ Lett.~ {\bf 83}, 3370 (1999);\\
L.~Randall and R.~Sundrum, Phys.~ Rev.~ Lett.~ {\bf 83}, 4690 (1999).
\end{thebibliography}
\end{document}